\documentclass[runningheads]{llncs}
\usepackage{graphicx}
\usepackage{amsmath}
\usepackage{amssymb}
\usepackage{array}
\usepackage{booktabs}
\usepackage{multirow}
\usepackage{hyperref}
\usepackage{siunitx}
\usepackage{makecell}
\usepackage{xcolor}
\begin{document}
%
\title{Lost in Tracking: Uncertainty-guided Cardiac Cine MRI Segmentation at Right Ventricle Base}
\titlerunning{Uncertainty-guided Cardiac Cine MRI Segmentation }
%
\author{Yidong Zhao \inst{1} \and Yi Zhang \inst{1} \and Orlando Simonetti \inst{2} \and Yuchi Han \inst{2} \and Qian Tao \inst{1}}
\institute{Department of Imaging Physics, Delft University of Technology, Lorentzweg 1, 2628 CJ Delft, The Netherlands\\ \email{q.tao@tudelft.nl} \and 
Cardiovascular Division, The Ohio State University Wexner Medical Center, Columbus, Ohio, USA
}
\authorrunning{Y. Zhao \textit{et al.}}

%
\maketitle              
\begin{abstract}
Accurate biventricular segmentation of cardiac magnetic resonance (CMR) cine images is essential for the clinical evaluation of heart function. However, compared to left ventricle (LV), right ventricle (RV) segmentation is still more challenging and less reproducible. Degenerate performance frequently occurs at the RV base, where the in-plane anatomical structures are complex (with atria, valve, and aorta) and vary due to the strong interplanar motion. In this work, we propose to address the currently unsolved issues in CMR segmentation, specifically at the RV base, with two strategies: first, we complemented the public resource by reannotating the RV base in the ACDC dataset, with refined delineation of the right ventricle outflow tract (RVOT), under the guidance of an expert cardiologist. Second, we proposed a novel dual encoder U-Net architecture that leverages temporal \emph{incoherence} to inform the segmentation when interplanar motions occur. The inter-planar motion is characterized by \emph{loss-of-tracking}, via Bayesian uncertainty of a motion-tracking model. Our experiments showed that our method significantly improved RV base segmentation taking into account temporal incoherence. Furthermore, we investigated the reproducibility of deep learning-based segmentation and showed that the combination of consistent annotation and loss of tracking could enhance the reproducibility of RV segmentation, potentially facilitating a large number of clinical studies focusing on RV\footnote{The refined RV annotation is accessible via \url{https://gitlab.tudelft.nl/yidongzhao/rvot_seg}.}. 
\end{abstract}
\section{Introduction}

Automatic segmentation of heart chambers is crucial for quantitatively assessing heart functions from cardiac magnetic resonance (CMR). Besides the left ventricle (LV), there is a growing clinical interest in accurate assessment of the right ventricle (RV), given its significance in heart and lung diseases~\cite{ho2006anatomy,han2021ranolazine,wang2020diagnostic,martin2023deep}. In recent years, several challenges have been dedicated to evaluating automatic CMR segmentation, including the ACDC challenge~\cite{bernard2018deep}, M\&M~\cite{campello2021multi}, and M\&M-v2~\cite{martin2023deep} which particularly focuses on RV. Results reveal that the nnUNet families~\cite{isensee2021nnu,arega2021using} have overall superior performance in biventricular segmentation, but RV has degenerated performance compared to that of LV~\cite{tao2019deep,bernard2018deep,campello2021multi,martin2023deep}. Degeneration is especially pronounced at the RV base, due to the irregular RV shape, large variability, and complex anatomical context~\cite{zhao2024artificial}. 
\begin{figure}[htb]
    \centering
    \includegraphics[width=1.0\textwidth]{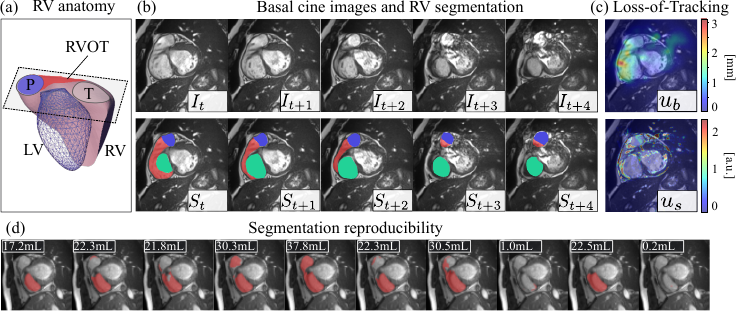}
    \caption{(a) Anatomy of LV and RV. The basal imaging plane covers the right ventricle outflow tract (RVOT), pulmonary valve (P), and tricuspid valve (T)~\cite{sheehan2008right}. (b) A short-axis basal slice contains atria (in green), P (in blue), and RVOT (in red), with complex and varying layouts. (c) Motion tracking has high uncertainty here ($u_b$ and $u_s$, defined in Section~\ref{sec:loss-of-tracking}), indicating loss-of-tracking. (d) RV segmentation by 10 Bayesian ensembles exhibits high uncertainty, resulting in a poorly reproducible volume estimation ranging from 0.2 to 37.8 mL.}
    \label{fig:teaser}
\end{figure}
Segmentation of basal slices is intrinsically challenging, because clinical cine MRI is 2D+$t$, with each plane imaged in a separate breath-hold, unable to capture the complex 3D+$t$ spatiotemporal motion at the base. Ventricles, atria, and valves all have inter-planar movements~\cite{martin2023deep,yilmaz2018evaluation}, as shown in Fig.~\ref{fig:teaser} (b). This complicates basal segmentation, resulting in quantitative errors of RV assessment~\cite{bernard2018deep,martin2023deep}.  

More specifically, RV segmentation error stems from the region of the right ventricle outflow tract (RVOT). RVOT is a pathway where the blood exits RV and enters the pulmonary artery~\cite{ho2006anatomy}, spanning from the right side of the tricuspid valve to the pulmonary valve~\cite{farre2011cardiac} (Fig.~\ref{fig:teaser} a). RVOT needs to be included for accurate RV quantification, but it is often overlooked in the annotations of public CMR datasets~\cite{yilmaz2018evaluation,bernard2018deep,campello2021multi,martin2023deep}. Common protocols delineate RV when the full cavity is covered~\cite{bernard2018deep,campello2021multi,martin2023deep} while RVOT is labeled as RV or background depending on cases or observers. This inconsistency in the annotation can affect the confidence of the neural network~\cite{kendall2017uncertainties}. Segmentation models, even when trained on the same dataset~\cite{gal2015dropout,lakshminarayanan2016simple,kendall2017uncertainties}, will have high uncertainty on basal slices, resulting in low \emph{reproducibility} of volume quantification. A typical example is shown in Fig.~\ref{fig:teaser} (d), where different segmentation models make varying RV predictions, indicating low reproducibility of RV volume estimation~\cite{gal2015dropout,zhao2022efficient,zhao2024bayesian}. This undermines the reliability of the assessment of RV function. 

Traditionally, temporal coherence is leveraged to improve the segmentation performance of CMR, because segmentation tends to be continuous in time and space~\cite{nilsson2016semantic,qin2018joint,yan2018left,yan2019cine,dong2020deu,bai2018recurrent}. Nilsson \textit{et al.} proposed a spatial-temporal Gated Recurrent Unit (ST-GRU)~\cite{nilsson2016semantic} to promote coherence of segmentation maps. In CMR segmentation, joint motion estimation and segmentation proved to be mutually beneficial~\cite{qin2018joint}. Yan \textit{et al.} proposed a flow-based feature fusion framework~\cite{yan2018left,yan2019cine} to integrate temporal coherence. Similarly, Wu \textit{et al.} explicitly encoded the flow as an additional feature for segmentation~\cite{wu2020cardiac,dong2020deu}. Bai \textit{et al.} leveraged recurrent neural network (RNN) for cine segmentation with registration-based pseudo-labels~\cite{bai2018recurrent}. 

However, the same principle does not apply to the RV base, because of the strong in-plane anatomy change (Fig.~\ref{fig:teaser} b). Intuitively, estimating the motion between temporal frames at RV base is ill-posed, i.e. a well-trained motion tracking model will fail to track due to the inter-planar motion, a phenomenon we hereafter call \textit{loss-of-tracking}. Hence, instead of leveraging the temporal coherence, we propose the opposite: we make use of the temporal \textit{incoherence}, which can be identified by motion tracking uncertainty. This uncertainty highlights the inter-planar motion of different structures (Fig.~\ref{fig:teaser} c) and is highly informative. 

We propose a novel loss-of-tracking-based method to tackle the currently unsolved RV base segmentation in CMR analysis, with the following contributions: 
\begin{itemize}
    \item For a more accurate RV definition, we complemented current public resources by providing refined RV base annotations for the ACDC dataset~\cite{bernard2018deep}, under the guidance of an expert cardiologist. This complemented community resource can be used to train and evaluate RV segmentation algorithms.
    \item We propose a Bayesian motion tracking framework for CMR cine, to estimate the tracking uncertainty (loss-of-tracking) which can identify the interplaner cardiac motion in an unsupervised manner. 
    \item We integrate this tracking uncertainty into a Dual-Encoder UNet architecture to enhance segmentation performance in the challenging regions of RV base.
    \item In addition, we demonstrated that the low reproducibility of deep learning segmentation can stem in part from the annotation inconsistency. Our work improved the RV segmentation reproducibility with refined RV annotation and loss-of-tracking. 
\end{itemize}

\section{Method}
\begin{figure}[htb]
    \centering
    \includegraphics[width=1.0\textwidth]{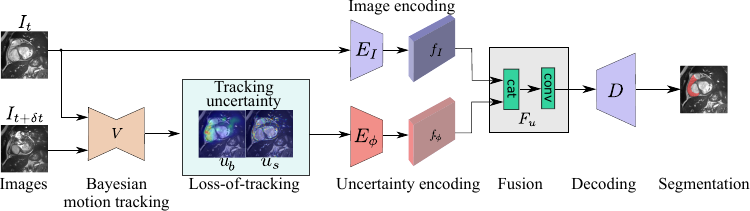}
    \caption{The Dual-Encoder UNet architecture for CMR segmentation: the upper path encodes the original image, while the lower path encodes the ``loss-of-tracking" from $I_t$ to $I_{t+\delta t}$, identified by a Bayesian motion-tracking model. The decoder $D$ predicts the segmentation map from the fused image feature $f_I$ and loss-of-tracking feature $f_{\phi}$.}
    \label{fig:method}
\end{figure}
\subsection{Loss-of-Tracking Detection via Registration Uncertainty}
\label{sec:loss-of-tracking}
We formulate the motion estimation problem as the registration between two temporal frames $I_t$ and $I_{t+\delta t}$ in a cine MRI. A VoxelMorph model~\cite{balakrishnan2019voxelmorph} $V$ is trained to predict the deformation field $\phi_t = V(I_t, I_{t+\delta t})$. Since the temporal cine images have similar contrast, we use the simple mean square error (MSE) $u_{s}$ as the image similarity metric, and regularize the smoothness of the estimated field $\phi_t$ via a gradient-based regularizer $R$. The loss function for motion tracking model training is:
\begin{align}
    \mathcal{L}_{reg}(\phi_t; I_t, I_{t+\delta t}) &= u_s(\phi_t\circ I_t, I_{t+\delta t}) + \lambda R(\phi_t)\\ \nonumber
    &= \Vert\phi_t \circ I_t - I_{t+\delta t} \Vert_2^2 + \lambda \Vert \nabla \phi_t \Vert_2^2,
\end{align}
where $\circ$ denotes the warping operation. At test time, the registration uncertainty can be inferred in two ways. First, we can evaluate the similarity between the warped image and the target image with $u_s$:
\begin{equation}
    u_s (\phi_t, I_t, I_{t+\delta t})= \Vert\phi_t \circ I_t - I_{t+\delta t} \Vert_2^2,
\end{equation}
where an elevated level of $u_s$ indicates regions of registration failure. Second, the model uncertainty of the trained registration network $V$ can be estimated from a Bayesian perspective~\cite{gal2015dropout,zhao2022efficient}, which derives the tracking uncertainty from the posterior $p(w|\mathcal{D})$ of network weights $w$, with $\mathcal{D}$ being the training dataset. The Bayesian uncertainty of motion $u_b$ is estimated via:
\begin{equation}
    u_b({\phi}_t, I_t, I_{t+\delta t}) \approx \mathsf{std} \left\{ V_{w_i}(I_t, I_{t+\delta t})\right\}_{i=1}^M,
\end{equation}
where $\mathsf{std}$ is the standard deviation operator, $\left\{w_i\sim p(w|\mathcal{D}) \right\}_{i=1}^{M}$ is a set of $M$ weights drawn from the posterior distribution, and $V_{w_i}$ denotes the trained model with weight $w_i$. We draw the posterior samples $\left\{w_i\sim p(w|\mathcal{D}) \right\}_{i=1}^{M}$ via the Hamiltonian Monte Carlo (HMC) method~\cite{chen2014stochastic,zhao2022efficient,zhao2024bayesian}.

\subsection{Uncertainty-guided Segmentation}
\subsubsection{Network Architecture} We propose a Dual-Encoder UNet architecture that takes both the image $I_t$ and its motion uncertainty $u_{\phi}$, which highlights the basal areas with interplanar motion, as input. Specifically, we use an encoder $E_I$ for image encoding and an additional encoder $E_\phi$ for loss-of-tracking encoding. The image encoder $E_I$ encodes the current temporal frame $I_t$ and outputs the feature $f_I = E_I(I_t)$. The loss-of-tracking encoder $E_\phi$ learns a representation $f_{\phi} = E_\phi (u_b, u_s)$ of the estimated motion uncertainty. The feature fusion is then performed by a learnable convolutional layer $F_u$, and the aggregated feature $f_a$ is expressed by
\begin{equation}
    f_a(I_t, u_b, u_s) = F_u\left( \mathsf{cat} \left(f_I, f_\phi\right) \right).
\end{equation}
Subsequently, $f_a$ is fed into the decoder $D$ for the final segmentation prediction. Skip connections between the encoder and decoder are preserved as in the original U-Net~\cite{ronneberger2015u}. The overall segmentation model is hence expressed as $\mathcal{S} = D\circ F_u \circ \mathsf{cat} \circ (E_I, E_{\phi})$. We show the proposed network architecture in Fig.~\ref{fig:method}.

\subsubsection{Bayesian Segmentation and Reproducibility} We used the same Bayesian HMC principle to generate a range of models~\cite{gal2015dropout,zhao2022efficient,zhao2024bayesian}. This Bayesian ensemble of segmentation networks is denoted by $\{\mathcal{S}_{\theta_j} \}_{j=1}^{M}$, parameterized by weight posterior samples $\left\{\theta_j\sim p(\theta|\mathcal{D}) \right\}_{j=1}^{M}$. Bayesian segmentation is performed via
\begin{equation}
    S_t = \frac{1}{M} \sum_{j=1}^M \mathcal{S}_{\theta_j}(I_t),\quad \mathrm{\sigma_v} (I_t) \approx \mathsf{std} \left\{ \mathcal{V}\circ {S}_{\theta_j}(I_t)\right\}_{j=1}^{M},
\end{equation}
where $S_t$ is the segmentation of $I_t$, $\mathcal{V}$ is the volume calculation operator, and $\sigma_v$ additionally quantifies the model reproducibility as the standard deviation (SD) of RV volume predicted by the Bayesian ensembles.

With ACDC standard labels, previous work has reported degenerate performance at the RV base, accompanied by high uncertainty~\cite{bernard2018deep,campello2021multi,martin2023deep,zhao2022efficient}. In the literature, degraded performance is often attributed to network generalizability or domain shift. However, a less explored hypothesis is that the training data can also play a role: if the annotation is inconsistent, the prediction from multiple trained models is also inconsistent. To validate the hypothesis and demonstrate the benefit of the refined RV annotation, we evaluated the RV segmentation reproducibility using the standard ACDC and our complemented ACDC annotation.
\section{Data and Experiments}
\subsubsection{Dataset}
We evaluated our method on the publicly available ACDC dataset~\cite{bernard2018deep}. It consists of cine images of 150 subjects, of which 100 serve as the training set, and the remaining 50 subjects are reserved for testing. Under the guidance of an expert cardiologist, we reannotated RV, including RVOT, on basal slices using the 3D Slicer~\cite{kikinis20133d} for all 150 subjects. In total, we manually refined the segmentation map on 240 slices of the original datasets (135 on the training split and 105 on the test split). In the following, we denote the original ACDC dataset as \textbf{Original}, and our relabeled dataset as \textbf{New}. We will open-source the new RV annotations on GitHub.

\subsubsection{Experimental Settings}
We used nnUNet~\cite{isensee2021nnu} as our segmentation backbone, sticking to its original loss function, optimizer, and network plan. Specifically, we trained a 2D nnUNet with all training samples, to be the baseline. To identify loss-of-tracking, we trained a motion tracking network using the VoxelMorph backbone~\cite{balakrishnan2019voxelmorph}, on images with a phase difference of $\delta t=4$.  Furthermore, we trained an {ST-GRU} network~\cite{nilsson2016semantic}, as a contrastive baseline that leverages temporal coherence for refined segmentation. For our Dual-Encoder UNet, we used the same encoder architecture for the loss-of-tracking input as for the image input, following the nnUNet design. We used $M=10$ HMC samples to build Bayesian ensembles, both for VoxelMorph (motion-tracking) and Dual-Encoder UNet (segmentation). 
\section{Results}
\subsection{Segmentation Accuracy}
We evaluated the segmentation performance of the three methods on RV, using the new label as ground truth. We divided each short-axis volume into basal, middle, and apical slices and evaluated segmentation accuracy on end-systolic (ES) and end-diastolic (ED) volumes separately. The accuracy measured by Dice coefficients is listed in Table~\ref{table:dice_results} on the three regions and the full volume. The table shows that the vanilla nnU-Net already forms a strong baseline for RV segmentation. Using the coherence-promoting ST-GRU leads to reduced segmentation accuracy. With the loss-of-tracking encoding, our proposed method outperforms the Vanilla U-Net, especially on the basal slices with an improvement of $1.2\%$ and $3.3\%$ at ED and ES volumes, respectively. In comparison, the improvement on the middle and apical slices is marginal in comparison with that on basal slices. 
\begin{table}[htb]
\caption{Segmentation accuracy of RV measured by the Dice coefficient [\%] on basal, middle, apical slices, and the full volume. Improvements with statistical significance ($p < 0.05$) using the Wilcoxon signed-rank test are labeled with *.}
\label{table:dice_results}
\centering
\begin{tabular}{p{1.9cm}p{1.2cm}<{\centering}p{1.2cm}<{\centering}p{1.2cm}<{\centering}p{1.2cm}<{\centering}p{1.2cm}<{\centering}p{1.2cm}<{\centering}p{1.2cm}<{\centering}p{1.2cm}<{\centering}  }
\toprule
\multirow{2}{*}{\textbf{Methods}} & \multicolumn{2}{c}{\textbf{Base}} & \multicolumn{2}{c}{\textbf{Mid}} & \multicolumn{2}{c}{\textbf{Apex}} & \multicolumn{2}{c}{\textbf{Full}} \\  \cmidrule(lr){2-3}\cmidrule(lr){4-5}\cmidrule(lr){6-7} \cmidrule(lr){8-9}
                         & ED  & ES  & ED  & ES & ED & ES & ED  & ES \\
\midrule
U-Net & \thead{90.1 \\ (±9.0)} & \thead{80.7 \\ (±22.4)} & \thead{94.5 \\ (±2.7)} & \thead{90.2 \\ (±5.1)} & \thead{86.9 \\ (±10.4)} & \thead{70.9 \\ (±26.1)} & \thead{92.9 \\ (±2.6)} & \thead{88.5 \\ (±4.7)}\\
ST-GRU & \thead{87.3\\ (±9.1)} & \thead{79.2\\ (±17.1)}  & \thead{91.0\\ (±4.2)} & \thead{87.0\\ (±5.6)}  & \thead{78.1\\ (±15.2)} & \thead{63.7\\ (±21.6)}  & \thead{89.4\\ (±3.4)} & \thead{84.9\\ (±4.9)} \\
Proposed & \thead{\textbf{91.3}{*} \\ (±7.3)} & \thead{\textbf{84.0}{*}  \\ (±16.5)} & \thead{\textbf{95.0}{*}  \\ (±2.6)} & \thead{\textbf{91.0}{*}  \\ (±5.0)} & \thead{\textbf{88.1}{*}  \\ (±9.8)} & \thead{\textbf{71.8}{*}  \\ (±26.8)} & \thead{\textbf{93.6} \\ (±2.3)} & \thead{\textbf{89.5}*  \\ (±4.5)}\\
\bottomrule
\end{tabular}
\end{table}

\begin{figure}[htb]
    \centering
    \includegraphics[width=\textwidth]{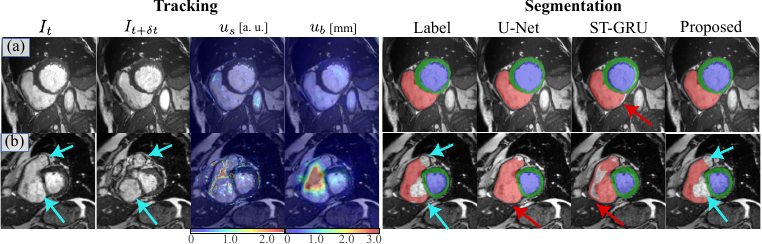}
    \caption{Qualitative results of tracking uncertainty and segmentation. The left panel shows the tracking uncertainty $u_b$ and $u_s$ between $I_t$ and $I_{t+\delta t}$. The right panel shows the segmentation labels and predictions. In case (b), the atrium and valve (cyan arrows) coexist with RV, and should not be included (c.f. the anatomy in Fig.~\ref{fig:teaser}).}
    \label{fig:qualitative-seg}
\end{figure}

In Fig.~\ref{fig:qualitative-seg}, we show some qualitative results of the loss-of-tracking detection and the predicted segmentation maps. In case (a), the right ventricle preserves its shape from  $I_t$ to $I_{t+\delta t}$, and the detected loss-of-tracking $u_s$ and $u_b$ stays on a relatively low level. On such images, all methods can correctly predict the RV segmentation map. However, the ST-GRU prediction still has a small deviation from the ground truth on ventricular borders (red arrow). We conjecture that ST-GRU suffers from imperfect motion tracking here.  In case (b), we show a basal slice with strong interplanar motion on which the RV and valves can hardly be distinguished from the single image $I_t$. The detected loss-of-tracking $u_b$ highlights the area that cannot be tracked from $I_t$ to $I_{t+\delta t}$, mainly in RVOT. The MSE pattern $u_s$ approximately delineates the separation between the valves and the RV. In this slice, the U-Net has difficulty in predicting RV segmentation in a single image $I_t$, but the proposed method can successfully predict the RV border with loss-of-tracking taken into account. In this case, taking segmentation consistency for granted like ST-GRU can harm the segmentation accuracy.

\subsection{Segmentation Reproducibility}
\begin{figure}[htb]
    \centering
    \includegraphics{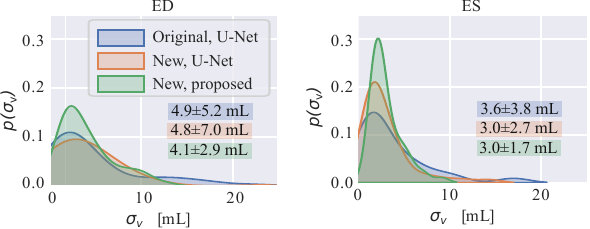}
    \caption{Distribution of segmentation reproducibility as measured by volume standard deviation $\sigma_v$. Statistics (mean ± std) are given in corresponding colors.}
    \label{fig:quantitative-sigma-v}
\end{figure}

In this section, we compare the reproducibility measured by the standard deviation of RV volume from Bayesian ensembles of segmentation networks. To validate the role of annotation, we repeated the experiments on both the Original and New ACDC annotations. Fig.~\ref{fig:quantitative-sigma-v} shows the distributions of $\sigma_v$ of the RV base in the ED and ES phases, respectively. The statistics are reported for the testing datasets.
\begin{figure}[htb]
    \centering
    \includegraphics{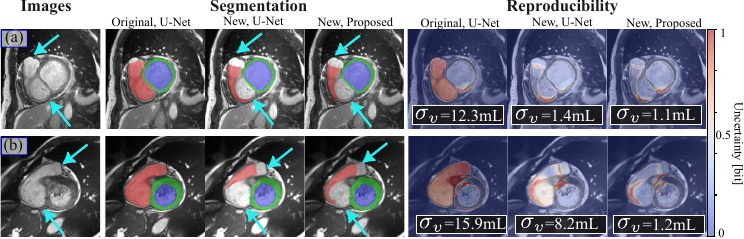}
    \caption{Examples of RVOT segmentation and reproducibility. High uncertainty indicates strong disagreement among different ensemble models. (a) Models trained with the original labels are uncertain on basal slices with both the valve and atria (cyan arrows) inplane. (b) The reproducibility is largely improved by the new annotations, and further reduced by the proposed method. }
    \label{fig:qualitative-reproducibility}
\end{figure}

All methods have a volume variance that peaked at a relatively low value ($< 5$ ml), especially at ES. However, we observe that the U-Nets trained with the original annotations have a longer tail than networks trained with the new annotations. The proposed method exhibits the highest reproducibility with a sharp peak. Fig.~\ref{fig:qualitative-reproducibility} (a) is a basal slice that covers the partial atrium and valve (cyan arrows) on the image plane. In this slice, we observe that the networks trained with the original annotations are highly uncertain, resulting in a volume SD of $12.3$ ml. In contrast, the networks trained with the new annotations successfully delineate the RVOT and have a reduced SD of $1.1\sim 1.4$ mL. In Fig.~\ref{fig:qualitative-reproducibility} (b), we show a case in which the vanilla U-Nets can have low reproducibility because the RVOT and atrium are not distinguishable, with $\sigma_v = 8.2$ mL. In comparison, our proposed method reduces $\sigma_v$ to $1.2$ mL. The results suggest that consistent annotation and loss-of-tracking can greatly improve reproducibility when segmenting difficult regions like the RV base.

\section{Conclusion}
Accurate biventricular segmentation of CMR cine images is important for the clinical evaluation of heart function. In this work, we set out to tackle the current challenges of segmenting RV base, for more accurate and reproducible RV assessment. We proposed a novel dual encoder U-Net architecture that leverages temporal incoherence, called \emph{loss-of-tracking}, to identify the interplanar motion at the base that previously deteriorated segmentation. Our experiments showed that \emph{loss-of-tracking} improved the segmentation of the RV base taking into account temporal incoherence. In addition, we complemented the public resource with refined RV base annotation including RVOT. Our work showed that the joint contribution of data and algorithm can lead to improved accuracy and reproducibility for the currently difficult regions of the RV base, potentially leading to more reliable RV assessment for future clinical studies.

\begin{credits}
\subsubsection{\ackname}
The authors gratefully acknowledge the TU Delft AI Initiative for
financial support. 
\subsubsection{\discintname}
The authors have no competing interests to declare that are
relevant to the content of this article.
\end{credits}

\clearpage 
\bibliographystyle{splncs04}
\bibliography{bibliography}
\end{document}